\documentclass[prd,preprintnumbers]{revtex4}

\usepackage{amsmath}
\usepackage{amsfonts}
\usepackage{graphicx}

\begin{document}

\tolerance=5000
\def\pp{{\, \mid \hskip -1.5mm =}}
\def\cL{{\cal L}}
\def\be{\begin{equation}}
\def\ee{\end{equation}}
\def\bea{\begin{eqnarray}}
\def\eea{\end{eqnarray}}
\def\tr{{\rm tr}\, }
\def\nn{\nonumber \\}
\def\e{{\rm e}}

\markboth{S. Nojiri and S.D. Odintsov}{Modified Gravity}

\title{INTRODUCTION TO MODIFIED GRAVITY AND GRAVITATIONAL ALTERNATIVE
FOR DARK ENERGY}

\author{Shin'ichi Nojiri\footnote{
Address from April: Dept. of Phys., Nagoya Univ., Nagoya 464-8602, Japan.
E-mail: snojiri@yukawa.kyoto-u.ac.jp, nojiri@cc.nda.ac.jp}}

\address{Department of Applied Physics,
National Defence Academy,
Hashirimizu Yokosuka 239-8686, Japan}

\author{Sergei D.~Odintsov\footnote{also at TSPU, Tomsk, E-mail:
odintsov@ieec.uab.es}}

\address{Instituci\`o Catalana de Recerca i Estudis
Avan\c{c}ats (ICREA)  and Institut de Ciencies de l'Espai
(IEEC-CSIC),
Campus UAB, Facultad Ciencies, Torre C5-Par-2a pl,
E-08193 Bellaterra (Barcelona), Spain}

\begin{abstract}

We review various modified gravities considered as gravitational
alternative for dark energy. Specifically, we consider the versions
of $f(R)$, $f(G)$ or $f(R,G)$ gravity, model with non-linear
gravitational coupling or string-inspired model
with Gauss-Bonnet-dilaton coupling in the late universe where they lead
to cosmic speed-up. It is shown that some of such theories may pass
the Solar System tests. On the same time, it is demonstrated that they
have quite rich cosmological structure: they may naturally describe
the effective (cosmological constant, quintessence or phantom) late-time era
with a possible transition from decceleration to acceleration thanks
to gravitational terms which increase with scalar curvature decrease.
The possibility to explain the coincidence problem as the manifestation of
the universe expansion in such models is mentioned.
The late (phantom or quintessence) universe filled with
dark fluid with inhomogeneous equation of
state (where inhomogeneous terms are originated from
the modified gravity) is also described.


\end{abstract}


\maketitle

\section{Introduction \label{S1}}

The dark energy problem (for recent review see \cite{copeland,p})
or, why current universe is expanding with the acceleration,
is considered to  be the one of the most fundamental theoretical problems
of XXI century. There are various directions aimed to construct the
acceptable dark energy model. Specifically, one can mention
scalar (quintessence or phantom) models, dark fluid with complicated
equation of state (EoS), more complicated field theories with fermions,
abelian or non-abelian vector field, string/M-theory, higher dimensions, etc.
Nevertheless, despite the number of attempts still there is no any
satisfactory explanation of dark energy origin. This is understandable
having in mind that even current values of cosmological parameters
are not yet defined with precise accuracy, and even less is known about
their evolution.

The modified gravity approach is extremely attractive in the applications
for late accelerating universe and dark energy. Indeed,

\noindent
1. Modified gravity provides the very natural gravitational alternative
for dark energy. The cosmic speed-up is explained simply by the fact of
the universe expansion where some sub-dominant terms (like $1/R$) may
become essential at small curvature.

\noindent
2. Modified gravity presents very natural unification of the early-time
inflation and late-time acceleration thanks to different role of
gravitational terms relevant at small and at large curvature.
Moreover, some models of modified gravity are predicted by string/M-theory
considerations.

\noindent
3. It may serve as the basis for unified explanation of dark energy and
dark matter. Some cosmological effects (like galaxies rotation curves)
may be explained in frames of modified gravity.

\noindent
4. Assuming that universe is entering the phantom phase,
modified gravity may naturally describe the transition from non-phantom
phase to phantom one without necessity to introduce the exotic matter
(like the scalar with wrong sign kinetic term or ideal fluid with EoS
parameter less than $-1$). In addition, often
the phantom phase in modified gravity is transient. Hence, no future Big
Rip is usually expected there.

\noindent
5. Modified gravity quite naturally describes the transition from
decceleration to acceleration in the universe evolution.

\noindent
6. The effective dark energy dominance may be assisted by the modification
of gravity. Hence, the coincidence problem is solved there simply by the
fact of the universe expansion.

\noindent
7. Modified gravity is expected to be useful in high energy physics (for
instance, for the explanation of hierarchy problem or unification of GUTs
with gravity).

\noindent
8. Despite quite stringent constraints from Solar System tests,
there are versions of modified gravity which may be viable theories
competing with General Relativity at current epoch. Nevertheless,
more serious check of such theories is necessary to fit them with various
observational data and precise Solar System tests.

These lectures are devoted to the review of some modified gravities
with above mentioned attractive features
and their cosmological properties at late universe. We do not discuss
here modified gravities related with brane-world approach as well as
Palatini formulation of modified gravities.

The paper is organized as follows. In sect.\ref{S2} we start from general $f(R)$
gravity in accelerating FRW universe. It is shown how this theory may be
rewritten as scalar-tensor gravity (Einstein frame). Two specific models:
the one with $1/R$ and $R^m$ terms, and another with $\ln R$ terms are
considered. It is shown that both models may lead to the (cosmological
constant or quintessence) acceleration of the universe as well as an early
time inflation. Moreover, the first model seems to pass the Solar System
tests, i.e. it has the acceptable newtonian limit, no instabilities and
no Brans-Dicke problem (decoupling of scalar) in scalar-tensor version.
The coupling of such modified gravity with usual matter is considered.
It is shown the occurence of transient phantom era in this case.
General conditions to $f(R)$ gravity admitting phantom divide crossing are
derived. Finally, it is demonstrated how scalar-tensor theory may be
rewritten as some version of modified gravity.
Sect.\ref{S3} is devoted to the study of FRW cosmology in string-inspired
dilaton gravity containing scalar-Gauss-Bonnet coupling. The occurence
of acceleration is again mentioned.
In sect.\ref{S4} the non-linear coupling of gravity with  matter is
considered. It is shown that some forms of such coupling may help to solve
the coincidence problem simply by the fact of the universe expansion.
  From another point, such modified gravity may also predict (cosmological
constant, quintessence or phantom) accelerating epoch. It is interesting
that the model with such non-linear coupling may serve for quite
interesting dynamical solution of cosmological constant problem.
The possible applications of modified gravity for solution of hierarchy
problem of high energy physics are described.

In sect.\ref{S5} we consider $f(G)$ and $f(R,G)$ theories where $G$ is
Gauss-Bonnet invariant. It is shown the possibility of cosmic speed-up
for various choices of function $f$, as well as possibility of
decceleration-acceleration transition. Sect.\ref{S6} is devoted to the study
of late-time cosmology for the universe filled with dark fluid with
inhomogeneous EoS. At first, the structure of  future singularities
is  classified for dark fluid with homogeneous EoS: $p=-\rho+A\rho^\alpha$.
Next, the impact of the inhomogeneous terms to future singularities
structure is analyzed for the same (effectively, phantom) dark fluid. As
the motivation for
inhomogeneous EoS dark fluid, the possible origin of the inhomogeneous
terms from scalar-tensor or modified gravity is indicated.
Some outlook and discussion are given in the last section.

\section{The modified $f(R)$ gravity \label{S2}}

In the present section $f(R)$ gravity in FRW universe is considered.
Special attention is paid to the versions of such theory
which may describe cosmic speed-up at late universe.
Let us start from the rather general 4-dimensional action:
\be
\label{M1}
S = \int d^4 x \sqrt{-g}\left\{ f(R) + L_m\right\}\ .
\ee
Here $R$ is the scalar curvature, $f(R)$ is an arbitrary function
and $L_m$ is a matter Lagrangian density.
The equation of the motion is given by
\be
\label{ANO1}
0=\frac{1}{2}g_{\mu\nu} f(R) - R_{\mu\nu}f'(R) - \nabla_\mu \nabla_\nu f'(R)
   - g_{\mu\nu}\nabla^2 f'(R) + \frac{1}{2} T_{\mu\nu}\ .
\ee
With no matter and for the Ricci tensor $R_{\mu\nu}$ being
covariantly constant, the equation of motion corresponding to the action
(\ref{M1}) is:
\be
\label{R19}
0=2f(R) - Rf'(R)\ ,
\ee
which is the algebraic equation with respect to $R$.
If the solution of Eq.(\ref{R19}) is positive, it expresses
deSitter
universe and if negative, anti-deSitter universe.
In the following,  the metric is assumed to be  in the FRW form:
\be
\label{RR33}
ds^2 = - dt^2 + \hat a(t)^2 \sum_{i=1}^3 \left(dx^i\right)^2\ .
\ee
Here we assume that the spatial part is flat as suggested by the
observation of
the Cosmic Microwave Background (CMB) radiation.
Without the matter and in FRW background, Eq.(\ref{ANO1}) gives
\be
\label{Cr4}
0=-\frac{1}{2}f(R) + 3\left(H^2  + \dot H\right) f'(R)
   -6\frac{\dot H}{H} f''(R) -18 H^2 \frac{d}{dt}\left(\frac{\dot H}{H^2}\right) f''(R)\ .
\ee
Here $R$ is given by $R=12H^2 + 6\dot H$.
Our main purpose is to look for accelerating cosmological solutions
of the following form:
de Sitter (dS) space, where $H$ is constant and $a(t)\propto \e^{Ht}$,
quintessence and phantom like cosmologies:
\be
\label{mGB2}
a=\left\{\begin{array}{ll} a_0t^{h_0},\ &\mbox{when}\ h_0>0\
\mbox{(quintessence)} \\
a_0\left(t_s - t\right)^{h_0},\ &\mbox{when}\ h_0<0\ \mbox{(phantom)} \\
\end{array} \right. \ .
\ee
The cosmologies of above type in various modified gravity models will
   be often discussed in this work.

Introducing the auxiliary fields, $A$ and
$B$, one can rewrite the action (\ref{M1}) as follows:
\be
\label{RR2b}
S=\int d^4 x \sqrt{-g} \left[{1 \over
\kappa^2}\left\{B\left(R-A\right) + f(A)\right\} + {\cal L}_{\rm
matter}\right]\ .
\ee
One is able to eliminate $B$,  and to obtain
\be
\label{RR6b}
S=\int d^4 x \sqrt{-g} \Bigl[{1 \over \kappa^2}\left\{f'(A)\left(R-A\right) + f(A)\right\}
+ {\cal L}_{\rm matter}\Bigr]\ ,
\ee
and by using the conformal transformation $g_{\mu\nu}\to \e^\sigma g_{\mu\nu}$
$\left(\sigma = -\ln f'(A)\right)$,
the action (\ref{RR6b}) is rewritten as the Einstein-frame action:
\be
\label{RR10}
S_E=\int d^4 x \sqrt{-g} \Bigl[{1 \over
\kappa^2}\left( R - {3 \over 2}g^{\rho\sigma}
\partial_\rho \sigma \partial_\sigma \sigma - V(\sigma)\right)
+ {\cal L}_{\rm matter}^{\sigma}\Bigr]\ .
\ee
Here,
\be
\label{RR11b}
V(\sigma) = \e^\sigma G\left(\e^{-\sigma}\right) -
\e^{2\sigma} f\left(G\left(\e^{-\sigma} \right)\right)
= {A \over f'(A)} - {f(A) \over f'(A)^2}\ .
\ee
The action  (\ref{RR6b}) is called the Jordan-frame action. In the
Einstein-frame action,
the matter couples with the scalar field $\sigma$.
Due to the scale transformation, there is a difference in time interval between the Einstein
and Jordan frames. Of course, we usually use clock to measure time. The clock
could be based essentially on  electromagnetism. In the Jordan frame, where the scalar field
$\sigma$ does not couple with matter, especially with electromagnetic
fields, the time measured by the clock is proportional to the cosmological time in the FRW metric.
On the other hand, in the Einstein frame where the scalar field $\sigma$ couples with matter,
the time measured by the clock is not always proportional to the cosmological time,
due to the coupling of the scalar field $\sigma$ with electromagnetic field.
The difference, of course, comes from the factor of the scale transformation, which includes
the scalar field $\sigma$. Since $\sigma$ is time-dependent in general,
the time measured by the clock becomes, say, slower than the cosmological
time in the Einstein frame. It is estimated the acceleration of the universe started
about five billion years ago. The five billion years are measured by using
the electromagnetism. Hence, these five billion years could be identified
with five billion years in the Jordan frame but not with those in the
Einstein frame (for further discussion of the relation between
Einstein and Jordan frames in cosmology, see review \cite{far}).

One may define the
effective EoS parameter $w_{\rm eff}$ in Jordan frame as
\be
\label{FRW3k}
w_{\rm eff}=\frac{p}{\rho}= -1 - \frac{2\dot H}{3H^2}\ ,
\ee
The scale factor in Einstein frame (when the two frames appear) is denoted
as $a(t)$.
Note that if we dont make transformation from one frame to another one,
then scale factor will be always denoted without hat.

\subsection{Modified gravity with negative and positive powers of the curvature \label{S21}}

As the first gravitational alternative for dark energy
we consider the following action \cite{SNPRD}
\be
\label{RR12}
f(R)= R - {c \over \left( R - \Lambda_1 \right)^n} + b \left( R -
\Lambda_2\right)^m \ .
\ee
Here we assume the coefficients $n$, $m$, $c$, $b>0$ but $n$, $m$ may be
fractional. Let us show that above model leads to acceptable cosmic
speed-up and is consistent with Solar System tests.

For the action (\ref{RR12}), Eq.(\ref{R19}) has the following form:
\be
\label{RR17}
0=-R + {(n+2)c \over \left(R-\Lambda_1\right)^n}
+ (m-2)b\left(R-\Lambda_2\right)^m\ .
\ee
Especially when $n=1$ and $m=2$, one gets
\be
\label{RR18}
R=R_\pm = {\Lambda_1 \pm \sqrt{\Lambda_1^2 +12c} \over 2}\ .
\ee
If $c>0$, one solution corresponds to deSitter space and another
to anti-deSitter.
If $-{\Lambda_1^2 \over 12}<c<0$ and $\Lambda_1>0$, both of solutions
express the
deSitter space. Hence, the natural possibility for the unification of
early-time inflation
with late-time acceleration appears (compare with proposal \cite{no}).
Here, higher derivative terms act in favour of early-time inflation while
$1/R$ term supports cosmic speed-up.

By assuming the FRW universe metric (\ref{RR33}), one may define the
Hubble rate by
$H=\dot{\hat a}/\hat a$.
   The contribution from matter may be neglected.
Especially when $n=1$, $m=2$, and $\Lambda_1=\Lambda_2=0$ in (\ref{RR12})
and the curvature is small, we obtain the following solution of
(\ref{ANO1}):
$\hat a \propto t^2$, which is consistent with the result in \cite{CDTT}.
If the present universe expands with above power law, the
curvature
of the present universe should be small compared with that of the deSitter universe
solution in (\ref{RR18}) with $\Lambda_1=0$. As the Hubble rate in the
present universe is $\left(10^{-33}{\rm eV}\right)^{-1}$, the parameter
$c$,
which corresponds to $\mu^4$ in \cite{CDTT}, should be much larger than
$\left(10^{-33}{\rm eV}\right)^4$.

We now consider the more general case that $f(R)$ is given by (\ref{RR12})
when the curvature is small.
Neglecting the contribution from the matter again, solving (\ref{ANO1}),
   we obtain
$\hat a \propto t^{(n+1)(2n+1) \over n+2}$.
It is quite remarkable that actually any negative power of the curvature
supports the cosmic acceleration. This gives the freedom in modification
of the model to achieve the consistency with  experimental tests of
newtonian gravity.

On the other hand, when the scalar curvature $R$ is large, one obtains
$\hat a \propto t^{-{(m-1)(2m-1) \over m-2}}$.
When ${(m-1)(2m-1) \over m-2}>0$, the universe is shrinking but if we change the arrow of time
by $t\to t_s - t$, the inflation occurs with the inverse power law and at $t=t_s$, the size of the universe
diverges. It is remarkable that when $m$ is fractional (or irrational) and $1<m<2$,
the expression of $\hat a$ is still valid and the power
becomes positive, the universe evolves with the (fractional) power law expansion.

It is interesting that the model of this section seems to be more
consistent than the one of ref. \cite{CDTT} (for discussion of newtonian
limit in $1/R$ theory see \cite{newton}) as it may pass Solar System
tests. Indeed,
in \cite{Dolgov}, small gravitational object like
the Earth or the Sun in the model \cite{CDTT} is considered. It has
been shown that the system  becomes instable.
This may cause the unacceptable force between distant Galaxies \cite{woodard}.
As shown in \cite{SNPRD}, however,
by adding the positive power (higher than 1)  of the scalar curvature
to the action, the instability could be significally improved.
Moreover, the account of quantum effects in modified gravity also acts
against instability \cite{mpla}.

It has been also mentioned in \cite{chiba} that $1/R$ model \cite{CDTT} which is
equivalent to some scalar-tensor gravity is ruled out as realistic theory
due to the constraints to such  Brans-Dicke type theories. In
\cite{SNPRD}, it was shown that by
adding the  scalar curvature squared term to the action, the mass
of the scalar field can be adjusted to be very large
   and the scalar field can decouple.
Hence, the modified gravity theory of this section (after some
fine-tuning \cite{SNPRD}) passes the Solar System tests.
In addition, at precisely the same values of parameters the above modified
gravity has the newtonian limit which does not deviate significally from
the one in  General Relativity.

\subsection{$\ln R$ gravity \label{S22}}

Other gravitational alternatives for dark energy may be suggested along the same line.
As an extension of the theory of the previous section, one may consider
the model \cite{ln} containing the logarithm of the scalar curvature $R$:
\be
\label{RD15}
f(R)=R + \alpha' \ln {R \over \mu^2} + \beta R^m\ .
\ee
We should note that  $m=2$ choice simplifies the model.
Assuming the scalar curvature is constant and the Ricci tensor is also
covariantly constant, the equations (\ref{R19}) are:
\be
\label{RD19}
0=2f(R) - Rf'(R) = \tilde f(R) \equiv R + 2\alpha' \ln {R \over \mu^2} - \alpha'\ .
\ee
If $\alpha'>0$, $\tilde f(R)$ is monotonically increasing function and
$\lim_{R\to 0}\tilde f(R)=-\infty$ and $\lim_{R\to +\infty}\tilde f(R)=+\infty$. There is
one and only one solution of (\ref{RD19}). This solution may correspond
to the inflation. On the other hand, if $\alpha'<0$, $\lim_{R\to 0}\tilde f(R)=\infty$
and $\lim_{R\to +\infty}=+\infty$. Since
$\tilde f'(R) = 1 + {\alpha' \over R}$, the minimum of $\tilde f(R)$, where $\tilde f'(R)=0$, is
given by $R=-2\alpha'$. If $\tilde f(-2\alpha')>0$, there is no solution of (\ref{RD19}).
If $\tilde f(-2\alpha')=0$, there is only one solution and if $\tilde f(-2\alpha')<0$,
there are two solutions.
Since $\tilde f(-2\alpha')=-2\alpha'\left(1 - \ln {-2\alpha' \over \mu^2}\right)$,
there are two solutions if $- {2\alpha' \over \mu^2}>\e$.
Since the square root of the curvature $R$ corresponds to the rate of the expansion
of the universe, the larger solution in two solutions might correspond to the
inflation in the early universe and the smaller one to the present accelerating
universe.

We can consider late FRW cosmology when the scalar curvature $R$ is small.
Solving (\ref{ANO1}), it follows that the power law inflation could occur:
$\hat a \propto t^{1 \over 2}$.
Since  $\dot{\hat a}>0$ but $\ddot{\hat a}<0$, the deccelerated expansion occurs.

One may discuss further generalizations like \cite{ln} (see also \cite{ln1})
\be
\label{GR1}
f(R)=R + \gamma R^{-n}\left(\ln {R \over \mu^2}\right)^m\ .
\ee
Here  $n$ is rectricted by $n>-1$ ($m$ is an arbitrary) in order that
the second term could be more dominant than the Einstein term
when $R$ is small.
For this model, we find
\be
\label{GR5}
\hat a\sim t^{(n+1)(2n+1) \over n+2}\ .
\ee
This does not depend on $m$. The logarithmic factor is almost irrelevant.
The effective $w_{\rm eff}$ is given by
\be
\label{GR6}
w_{\rm eff}=-{6n^2 + 7n - 1 \over 3(n+1)(2n+1)}\ .
\ee
Then $w_{\rm eff}$ can be negative if
\be
\label{GR7}
-1<n<-{1 \over 2}\ \mbox{or}\
n>{-7 + \sqrt{73} \over 12}=0.1287\cdots \ .
\ee
   From (\ref{GR5}), the condition that the universe
could accelerate is ${(n+1)(2n+1) \over n+2}>1$, that is:
\be
\label{GR8}
n> {-1 + \sqrt{3} \over 2}=0.366\cdots \ .
\ee
Clearly, the effective EoS parameter  $w$ may be within the existing
bounds.

\subsection{Modified gravity coupled with matter \label{S23}}

In this subsection we will show that the modified gravity may lead to the
   effective  phantom or quintessence dark energy
without necessity to introduce the (negative kinetic energy) phantom scalar
   or negative pressure ideal fluid.
In fact, the matter is taken to be the usual ideal fluid.

The ideal fluid is taken as the matter with the constant $w$: $p=w\rho$.
Then from the energy conservation law it follows
$\rho = \rho_0 a^{-3(1+w)}$ .
In a some limit, strong cuvature or weak one, $f(R)$ may behave as
$f(R)\sim f_0 R^\alpha$, with constant $f_0$ and $\alpha$.
An exact solution of the equation of motion is found to be \cite{ANO}
\bea
\label{M8}
&& a=a_0 t^{h_0} \ ,\quad h_0\equiv \frac{2\alpha}{3(1+w)} \ ,\nn
&& a_0\equiv \left[-\frac{6f_0h_0}{\rho_0}\left(-6h_0 + 12 h_0^2\right)^{\alpha-1}
\left\{\left(1-2\alpha\right)\left(1-\alpha\right) - (2-\alpha)h_0\right\}\right]^{-\frac{1}{3(1+w)}}\ .
\eea
When $\alpha=1$, the result $h_0 = \frac{2}{3(1+w)}$ in the Einstein gravity is reproduced.
The effective $w_{\rm eff}$ may be defined by $h_0=\frac{2}{3\left(1+w_{\rm eff}\right)}$.
By using (\ref{M8}), one finds the effective $w_{\rm eff}$  (\ref{FRW3k}) is given by
\be
\label{M9}
w_{\rm eff}=-1 + \frac{1+w}{\alpha}\ .
\ee
Hence, if $w$ is greater than $-1$ (effective quintessence or even usual ideal fluid with positive $w$),
when $\alpha$ is negative, we obtain the effective phantom phase where $w_{\rm eff}$ is less than $-1$.
This is different from the case of pure modified gravity.
Moreover, when $\alpha>w+1$ (it can be even positive),  $w_{\rm eff}$ could be negative (for negative $w$).
Hence, it follows that modified gravity minimally coupled with usual (or quintessence) matter may reproduce
quintessence (or phantom) evolution phase for dark energy universe in an easier way than without such coupling.

One may now take $f(R)$ as
\be
\label{BRXXX1}
f(R)={1 \over \kappa^2} \left(R - \gamma R^{-n} + \eta R^2\right)\ .
\ee
When the cuvature is small, the second term becomes dominant and one may identify
$f_0=-\frac{\gamma}{\kappa^2}$ and $\alpha=-n$. Then from (\ref{M9}),
it follows $w_{\rm eff}=-1 - \frac{1+w}{n}$.
Hence, if $n>0$,  an effective phantom era occurs even if $w>-1$. Usually
the phantom era generates the future Big Rip singularity \cite{brett}.
(Note that there may occur  different types of future singularities,
say, sudden ones \cite{Barrow,nojiri,tsujikawa,Stefancic}, etc. The
complete
classification of Big
Rip singularities is given in \cite{tsujikawa}.)
However, near the Big Rip singularity,
the curvature becomes large and the last term  becomes dominant. In this case $\alpha=2$ and
$w_{\rm eff}=\frac{-1+w}{2}$. Then if $w>-1$, it follows $w_{\rm eff}>-1$,
which prevents the Big Rip singularity (it makes the phantom phase transient).
To conclude, it looks quite promising that modest modification of General Relativity coupled
to ideal fluid matter leads to an effective dark energy universe in the
very natural way.

\subsection{The equivalence with scalar-tensor theory \label{S24}}

It is very interesting that
$f(R)$ gravity is in some sense equivalent to the scalar-tensor theory
with the action:
\bea
\label{k1}
S=\int d^4 x \sqrt{-g}\left\{\frac{1}{2\kappa^2}R
   - \frac{1}{2}\omega(\phi)\partial_\mu \phi\partial^\mu \phi
   - V(\phi)\right\}\ , \nn
\omega(\phi)=- \frac{2}{\kappa^2}h'(\phi)\ ,\quad
V(\phi)=\frac{1}{\kappa^2}\left(3h(\phi)^2 + h'(\phi)\right)\ .
\eea
Here $h(\phi)$ is a proper function of the scalar field $\phi$.
Imagine  the following FRW cosmology is constructed:
\be
\label{k7}
\phi=t\ ,\quad H=h(t)\ .
\ee
Then , as it has been demonstrated in ref.\cite{e}, {\it any} cosmology
defined by $H=h(t)$ in (\ref{k7}) can be realized by (\ref{k1}).
Indeed, if one defines a new field $\varphi$ as
\be
\label{f5}
\varphi=\int d\phi \sqrt{\left|\omega(\phi)\right|}\ ,
\ee
the action (\ref{k1}) can be rewritten as
\be
\label{f6}
S=\int d^4 x
\sqrt{-g}\left\{\frac{1}{2\kappa^2}R \mp \frac{1}{2}\partial_\mu
\varphi \partial^\mu \varphi  - \tilde V(\varphi)\right\}\ .
\ee
The sign in front of the kinetic term depends on the sign of
$\omega(\phi)$. If the sign of $\omega$ and therefore the sign of
$\dot H$ is positive (negative), the sign of the kinetic term is
$-$ ($+$). Therefore, in the phantom phase, the sign is always $+$
and in the non-phantom phase, always $-$. One assumes $\phi$ can
be solved with respect to $\varphi$: $\phi=\phi(\varphi)$. Then
the potential $\tilde V(\varphi)$ is given by
$\tilde V(\varphi)\equiv V\left(\phi(\varphi)\right)$.
Since $\tilde V(\varphi)$ could be uniquely determined, there is one-to-one
correspondence between $H$ and $\tilde V(\varphi)$.

In case the sign in front of the kinetic term of
$\varphi$ in (\ref{f6}) is $-$, we can use the conformal
transformation $g_{\mu\nu}\to \e^{\pm\kappa \varphi\sqrt{\frac{2}{3}}}g_{\mu\nu}$,
and make the kinetic term of $\varphi$ vanish. Hence, one obtains
\be
\label{f8}
S=\int d^4 x \sqrt{-g}\left\{\frac{\e^{\pm\kappa\varphi\sqrt{\frac{2}{3}}}}{2\kappa^2}R
   - {\e^{\pm 2\kappa \varphi\sqrt{\frac{2}{3}}}}\tilde V(\varphi)\right\}\ .
\ee
The action  (\ref{f8}) may be called as Jordan frame action
and the action (\ref{f6}) as the Einstein frame action. Since
$\varphi$ becomes the auxiliary field, one may delete $\varphi$ by
using an equation of motion:
\be
\label{f9}
R=\e^{\pm\kappa \varphi\sqrt{\frac{2}{3}}}
\left(4\kappa^2 \tilde V(\varphi) \pm 2\kappa \sqrt{\frac{3}{2}}
\tilde V'(\varphi)\right)\ ,
\ee
which may be solved with respect to $R$ as $\varphi=\varphi(R)$.
One can rewrite the action (\ref{f8}) in the form of $f(R)$ gravity \cite{e}:
\bea
\label{f10}
S&=&\int d^4 x \sqrt{-g}f(R)\ , \nn
f(R) &\equiv&
\frac{\e^{\pm \kappa \varphi(R)\sqrt{\frac{2}{3}}}}{2\kappa^2}R
    - {\e^{\pm 2\kappa \varphi(R)\sqrt{\frac{2}{3}}}}\tilde V\left(\varphi(R)\right)\ .
\eea
Note that one can rewrite the scalar-tensor theory (\ref{k1})
or equivalently (\ref{f6}), only when the sign in front of the
kinetic term is $-$ in (\ref{f6}), that is, $\omega(\phi)$ is
positive. In the phantom phase, $\omega(\phi)$ is negative. In
this case, the above trick to rewrite the phantom scalar-tensor
theory as  $f(R)$ gravity does not work. We should note, however,
as the metric is conformally transformed, even if the universe
in the original metric (corresponding to the Einstein frame) is in
the phantom phase, the universe in the rescaled metric
(corresponding to the Jordan frame) can be in non-phantom phase in
general. In addition, one should bear in mind that the same FRW dynamics
(same scale factor) may be obtained from one of these two mathematically
equivalent descriptions. Nevertheless, the properties of such FRW
cosmology in both models may be quite different (for instance, more FRW
solutions may exist, say, for scalar-tensor theory, newton law may be
different, stability of such equivalent solutions may be different, etc).

\subsection{Transition from phantom  to non-phantom phase \label{S25}}

Assuming that our universe is entering into the phantom phase recently,
and having in mind the occurence of (cosmologically recent)
decceleration-acceleration transition,  the question appears:
what are the conditions for phantom-non-phantom transition in modified
$f(R)$ gravity? For instance, the conditions for the existence of
deSitter universe solution in $f(R)$ gravity may be derived in the form of
algebraic
equation \cite{cognola1}. The stability of deSitter universe solution was
established in ref.\cite{cognola1} where one-loop quantization of $f(R)$
was done. (The same stability criteria was also found in
ref.\cite{faraoni} using perturbations analysis.)

The effective EoS parameter is defined by (\ref{FRW3k}).
We now define a time-dependent parameter $F(t)$ by
\be
\label{Cr2}
F(t)\equiv - \frac{2\dot H}{3H^2}\ ,
\ee
that is, $F(t)=1+w_{\rm eff}$.
Note that  $F=0$ at the transition point between non-phantom
($w_{\rm eff}>-1$ or $F>0$)
and phantom ($w_{\rm eff}<-1$ or $F<0$) phases.

We now consider the region near the transition point between non-phantom and phantom phases
and assume $F$ is small.
Then Eq.(\ref{Cr4}) reduces to
\be
\label{Cr5}
0=-\frac{1}{2}f(12H^2) + 3H^2 f'(12H^2)
   + 81 H^4 F f''(12H^2) + 27 H^3 \dot F f''(12H^2) + {\cal O}\left(F^2\right)\ .
\ee
If the transition point corresponds to the deSitter solution, which satisfies
\be
\label{Cr6}
0=-\frac{1}{2}f(12H^2) + 3H^2 f'(12H^2) \ ,
\ee
one finds $\dot F\sim -3 H F$. Then $F$ behaves as $F\sim F_0 \e^{-3Ht}$,
which does not vanish in a finite time and therefore there does not occur
the transition.
On the other hand, if the transition point does NOT correspond to the deSitter solution,
it follows
\be
\label{Cr9}
\dot F \sim \frac{f(12H^2) - 6H^2 f'(12H^2)}{54 H^3 f''(12H^2)}\ .
\ee
Then near the transition point if
\bea
\label{Cr10}
&& f''(12H^2)>0\ \mbox{and}\ f(12H^2) - 6H^2 f'(12H^2)<0 \nn
\mbox{or} && f''(12H^2)<0\ \mbox{and}\ f(12H^2) - 6H^2 f'(12H^2)>0 \ ,
\eea
there could occur the transition from the non-phantom phase ($F>0$) to phantom phase ($F<0$).

As an example, we consider
\be
\label{CrEx1}
f(R)=f_0\e^{R/6H_0^2}\ ,
\ee
with constants $f_0$ and $H_0$ (for similar choice with negative $R$ in
the exponent see \cite{dadhich}.) From (\ref{Cr9}), one finds
\be
\label{CrEx2}
\dot F \sim \frac{2H_0^2\left(H_0^2 - H^2\right)}{3H^3}\ .
\ee
Then by properly choosing initial conditions, if $H>H_0$ when $F\sim 0$ but positive
(that is, in non-phantom phase), we find $\dot F<0$ and there could occur the transition
from non-phantom phase to phantom phase (crossing of phantom divide).

\section{String-inspired Gauss-Bonnet gravity as dark energy \label{S3}}

The successful dark energy theory may be searched in string/M-theory.
Indeed, it is quite possible that some unusual gravity-matter
couplings predicted by the fundamental theory may become important at
current, low-curvature universe (being not essential in intermediate epoch
from strong to low curvature). For instance, in the study of
string-induced gravity near to initial singularity the role of Gauss-Bonnet (GB)
coupling with scalar was quite important for
ocurrence of non-singular cosmology \cite{ART,nick}.
Moreover, string/M-theory naturally predicts the appearence of the terms
with inverse powers of curvature invariants in compactified, low-energy
effective action \cite{stringplb}.

The present section is devoted to the study
of the role of GB coupling with the scalar field  (which naturally
appears in low-energy effective action) to the late-time
universe \cite{NOS}. It is explicitly demonstrated that such term itself
cannot induce
the effective phantom late-time universe if the scalar is canonical in the
absence of potential term. It may produce the
effective quintessence (or phantom) era, explaining the current acceleration
only when the scalar is phantom or when the scalar is canonical with non-zero
potential. It is interesting that it may
also have the important impact to the Big Rip singularity, similarly
to quantum effects \cite{nojiri,final,tsujikawa}, preventing it. Note
that we concentrate
mainly on the exponential scalar-GB coupling and exponential scalar
potential, while the consideration of other types of such functions and
their role in late time cosmology may be considered too \cite{neupane}.

We consider a model of the scalar field $\phi$ coupled with gravity.
As a stringy correction, the term proportional to the GB
invariant $G=R^2 - 4 R_{\mu\nu} R^{\mu\nu} + R_{\mu\nu\rho\sigma} R^{\mu\nu\rho\sigma}$ is added.
The starting action is given by
\bea
\label{GB2}
&& S=\int d^4x \sqrt{-g}\left\{ \frac{1}{2\kappa^2}R
      - \frac{\gamma}{2}\partial_\mu \phi \partial^\mu \phi
      - V(\phi) + f(\phi) G\right\}\ ,\nn
&& V=V_0\e^{-\frac{2\phi}{\phi_0}}\ ,\quad f(\phi)=f_0 \e^{\frac{2\phi}{\phi_0}}\ .
\eea
Here $\gamma=\pm 1$. For the canonical scalar, $\gamma=1$ but at least
when GB term is not included, the scalar behaves as phantom only
when $\gamma=-1$.

Starting with FRW universe metric (\ref{FRW3k}) in sect.\ref{S2} and assuming (\ref{mGB2}) in sect.\ref{S2},
the following solutions may be obtained
\bea
\label{GB15}
V_0 t_1^2&=& - \frac{1}{\kappa^2\left(1 + h_0\right)}\left\{3h_0^2 \left( 1 -
h_0\right)
+ \frac{ \gamma \phi_0^2 \kappa^2 \left( 1 - 5 h_0\right)}{2}\right\}\ ,\nn
\frac{48 f_0 h_0^2}{t_1^2}&=& - \frac{6}{\kappa^2\left( 1 +
h_0\right)}\left(h_0
      - \frac{\gamma \phi_0^2 \kappa^2}{2}\right)\ .
\eea
Even if $\gamma=-1$, there appear the solutions describing non-phantom
cosmology coresponding the quintessence or matter.
As an example, we consider the case that $h_0=-\frac{80}{3}<-1$, which gives
$w_{\rm eff}= - 1.025$. Simple tuning gives other acceptable values of the
effective $w$ in the range close to $-1$.
This is consistent with the observational bounds for effective $w$ (for recent
discussion and complete list of refs., see \cite{lazkoz}).
Then from (\ref{GB15}), one obtains
\be
\label{GR22}
V_0t_1^2 = \frac{1}{\kappa^2}\left( \frac{531200}{231}
+ \frac{403}{154}\gamma \phi_0 \kappa^2 \right)\ ,\quad
\frac{f_0}{t_1^2} = -\frac{1}{\kappa^2}\left( \frac{9}{49280}
+ \frac{27}{7884800}\gamma \phi_0 \kappa^2 \right)\ .
\ee
Therefore even starting from the canonical scalar theory with positive
potential, we may
obtain a solution which reproduces the observed value of $w$.

There is another kind of solutions \cite{NOS}. If $\phi$ and
$H$ are constants:
$\phi=\varphi_0$, $H=H_0$, this corresponds to deSitter space.
Then the solution of equations of motion gives:
\be
\label{GB29}
H_0^2 = - \frac{\e^{-\frac{2\varphi_0}{\phi_0}}}{8f_0 \kappa^2}\ .
\ee
Therefore in order for  the solution to exist, the condition is $f_0<0$.
In
(\ref{GB29}), $\varphi_0$ can be arbitrary. Hence, the Hubble rate $H=H_0$
might be determined by an initial condition.

In \cite{NOS}, the stability of the above  (and related) solutions
was checked and it was found that
the case corresponding to phantom cosmology with $h_0<0$ is always stable
but the case corresponding to non-phantom cosmology with $h_0>0$ is always unstable.
Also in \cite{stringT}, by taking into account the higher-order
string corrections to Einstein-Hilbert action, the evolution of (phantom) dark energy universe
was studied. When fixed dilaton, while the presence of a cosmological constant gives stable
de-Sitter fixed points in the cases of heterotic and bosonic strings, no stable de-Sitter
solutions exist when a phantom fluid is present.
It was found that the universe can exhibit a Big Crunch singularity with a finite time for
type II string, whereas it reaches a Big Rip singularity for heterotic and bosonic strings.
Thus the fate of dark energy universe crucially depends upon the type of string theory
under consideration.
Furthermore, a barotropic perfect fluid coupled to the scalar field
(dilaton or
compactification modulus) was also considered and phase space analysis and the stability
of asymptotic solutions are performed for a number of models which include
($i$) fixed scalar field,
($ii$) linear dilaton in string frame, and
($iii$) logarithmic modulus in Einstein frame.
It was also shown that dilaton Gauss-Bonnet gravity (with no matter)
   may not explain the current
acceleration of the universe. It was also studied the
future evolution of the universe using the GB parametrization and found
that Big Rip singularities
can be avoided even in the presence of a phantom fluid because of the balance between the fluid
and curvature corrections. A non-minimal coupling between the fluid and the modulus field also
opens up the interesting possibility to avoid Big Rip regardless of the
details of the fluid
equation of state. Note also that above action has interesting
applications in early time string cosmology (for recent discussion and
list of references, see \cite{rb}).
Definitely, dilaton Gauss-Bonnet gravity (perhaps, also with account of
higher order terms like Euler invariant) deserves very deep investigation
as possible candidate for dark energy.

\section{Modified gravity: non-linear coupling, cosmic acceleration and
hierarchy problem \label{S4}}

\subsection{Gravitational solution of coincidence problem \label{S41}}

It is interesting to investigate the role of non-minimal coupling of
modified gravity with dark energy to cosmic acceleration epoch.
The corresponding, quite general theory has been suggested in
ref.\cite{coupled} (see also \cite{hjs}). As an example of such
theory,
the following action is considered:
\be
\label{LR1}
S=\int d^4 x \sqrt{-g}\left\{{1 \over \kappa^2}R + \left(\frac{R}{\mu^2}\right)^\alpha L_d \right\}\ .
\ee
Here $L_d$ is matter-like action (dark energy). The choice of  parameter
$\mu$ may keep away the unwanted instabilities which often occur in higher
derivative theories.
The second term in above action describes the non-linear coupling of
matter with gravity (in parallel with  $R \phi^2$ term which is
usually required by renormalizability condition \cite{buchbinder}).
Similarly, such term may be induced by quantum effects as non-local effective action. Hence, it
is natural to consider that it belongs to matter sector. (Standard matter
is not included for simplicity).

By the variation over $g_{\mu\nu}$, the equation of motion follows:
\be
\label{LR2}
0= {1 \over \sqrt{-g}}{\delta S \over \delta g_{\mu\nu}}
= {1 \over \kappa^2}\left\{{1 \over 2}g^{\mu\nu}R - R^{\mu\nu}\right\}
+ \tilde T^{\mu\nu}\ .
\ee
Here the effective
EMT tensor  $\tilde T_{\mu\nu}$ is defined by
\bea
\label{w5}
\tilde T^{\mu\nu}&\equiv& \frac{1}{\mu^{2\alpha}}\left\{ - \alpha R^{\alpha - 1} R^{\mu\nu} L_d
+ \alpha\left(\nabla^\mu \nabla^\nu
   - g^{\mu\nu}\nabla^2 \right)\left(R^{\alpha -1 } L_d\right) + R^\alpha T^{\mu\nu}\right\}\ ,\nn
T^{\mu\nu}&\equiv& {1 \over \sqrt{-g}}{\delta \over \delta g_{\mu\nu}}
\left(\int d^4x\sqrt{-g} L_d\right)
\eea
Let free massless scalar be a matter
\be
\label{LR4}
L_d = - {1 \over 2}g^{\mu\nu}\partial_\mu \phi \partial_\nu \phi\ .
\ee
Then the equation given by the variation over $\phi$ has the following form:
\be
\label{LR5}
0= {1 \over \sqrt{-g}}{\delta S \over \delta \phi}={1 \over
\sqrt{-g}}\partial_\mu
\left(R^\alpha \sqrt{-g} g^{\mu\nu}\partial_\nu \phi \right)\ .
\ee
The metric again corresponds to FRW universe with flat 3-space.
If we assume $\phi$ only depends on $t$ $\left(\phi=\phi(t)\right)$, the solution of
scalar field equation (\ref{LR5}) is given by
\be
\label{LR9}
\dot \phi = q a^{-3} R^{-\alpha}\ .
\ee
Here $q$ is a constant of the integration. Hence $R^\alpha L_d = {q^2 \over 2 a^6 R^\alpha}$,
which becomes dominant when $R$ is small (large) compared with the
Einstein term ${1 \over \kappa^2}R$ if $\alpha>-1$ $\left(\alpha <-1\right)$.
Thus, one arrives at the remarkable possibility \cite{coupled} that dark
energy grows
to asymptotic dominance over the usual matter with decrease of the curvature.
At current universe, this solves the coincidence problem (the equality of
the energy density for dark energy and for matter) simply by the fact of
the universe expansion.

Substituting (\ref{LR9}) into (\ref{LR2}),  the $(\mu,\nu)=(t,t)$
component of equation of motion has the following form:
\bea
\label{LR11}
0&=&-{3 \over \kappa^2}H^2 + \frac{36q^2}{\mu^{2\alpha} a^6\left(6\dot H + 12 H^2\right)^{\alpha + 2}}
\left\{ {\alpha (\alpha + 1) \over 4}\ddot H H + {\alpha + 1 \over 4}{\dot H}^2 \right. \nn
&& \left. + \left(1 + {13 \over 4}\alpha + \alpha^2\right)\dot H H^2
+ \left(1 + {7 \over 2}\alpha\right) H^4 \right\}\ .
\eea
The accelerated FRW solution of (\ref{LR11}) exists \cite{coupled}
\be
\label{LR13}
a=a_0 t^{\alpha + 1 \over 3}\quad \left(H={\alpha + 1 \over 3t}\right)\ ,\quad
a_0^6 \equiv \frac{\kappa^2 q^2 \left(2\alpha - 1\right)\left(\alpha - 1\right) }
{\mu^{2\alpha}  3\left(\alpha + 1\right)^{\alpha + 1}
\left({2 \over 3}\left(2\alpha - 1\right)\right)^{\alpha + 2}}\ .
\ee
Eq.(\ref{LR13}) tells that the universe accelerates, that is, $\ddot a>0$
if $\alpha>2$. If $\alpha<-1$, the solution (\ref{LR13}) describes shrinking universe if $t>0$.
If the time is shifted as $t\to t - t_s$ ($t_s$ is a constant), the
accelerating and expanding
universe occurs when $t<t_s$. In the solution with $\alpha<-1$ there
appears a Big Rip singularity at $t=t_s$.
For the matter with the relation $p=w\rho$, where $p$ is the pressure and $\rho$ is the
energy density, from the usual FRW equation, one has $a\propto t^{2 \over 3(w+1)}$.
For $a\propto t^{h_0}$ it follows $w=-1 + {2 \over 3h_0}$,
and the accelerating expansion ($h_0>1$) of the universe occurs if
$-1<w<-{1 \over 3}$.
For the case of (\ref{LR13}), one finds
\be
\label{LRa2}
w={1 - \alpha \over 1 + \alpha}\ .
\ee
Then if $\alpha<-1$, we have $w<-1$, which is an effective phantom.
For the general matter with the relation $p=w\rho$ with constant $w$,
the energy $E$ and the energy density $\rho$ behave as
$E\sim a^{-3w}$ and $\rho\sim a^{-3\left(w + 1\right)}$.
Thus, for the standard phantom with $w<-1$, the density becomes large with
time and might generate the Big Rip.
Hence, it is demonstrated that non-linear coupling of gravity with matter
may produce the accelerated cosmology with the effective EoS parameter
being close (above, equal or below) to $-1$. More complicated couplings
\cite{coupled} may be considered in the same fashion.

\subsection{Dynamical cosmological constant theory: an exact example \label{S42}}

There are many proposals to solve the cosmological constant problem
dynamically.
One of the possible solutions of the cosmological constant
problem is pointed out by  Mukohyama and  Randall \cite{MR} (see also
\cite{Dolgov2}).
In \cite{MR}, the following  action similar to the one under
consideration has been proposed:
\begin{equation}
\label{MR1}
I=\int d^4 x\sqrt{-g}\left[\frac{R}{2\kappa^2} + \alpha_0 R^2 +
\frac{\left(\kappa^4\partial_\mu \varphi
\partial^\mu \varphi\right)^q}{2q \kappa^4 f(R)^{2q-1}} - V(\varphi)\right]\ ,
\end{equation}
where $f(R)$ is a proper function. When the
curvature is small, it is assumed
$f(R)$ behaves as
\begin{equation}
\label{MRIO1}
f(R)\sim \left(\kappa^2 R^2 \right)^m\ .
\end{equation}
Here  $m$ is positive.
When the curvature is small, the vacuum energy, and therefore the value of the
potential becomes small.
Then one may assume, for the small curvature, $V(\varphi)$ behaves as
\begin{equation}
\label{MRIO2}
V(\varphi)\sim V_0 \left(\varphi - \varphi_c\right)\ .
\end{equation}
Here $V_0$ and $\varphi_c$ are constants. If $q>1/2$, the factor in front of
the kinetic term of
$\varphi$ in (\ref{MR1}) becomes large. This makes the time development of the
scalar field $\varphi$
very slow and it is expected that $\varphi$ does not reach $\varphi_c$. This
model may explain the
acceleration of the present universe. The model  (\ref{MR1}) is also expected
to be stable
under the radiative corrections. In fact, when the curvature is small and the
time development of the
curvature can be neglected, if we rescale the scalar field $\varphi$ as
$\varphi\to R^{m(2q-1)/q}\varphi$, the curvature in the kinetic term can be
absorbed into the
redefinition of $\varphi$ and there appear factors including $R^{m(2q-1)/q}$.
Therefore
if $m(2q-1)/q>0$, the interactions could be suppressed when the curvature is
small and there will not
appear the radiative correction to the vacuum energy except the one-loop
corrections.

There is an exactly solvable model \cite{inagaki}, which realizes the
above scenario.
Let us choose
\begin{equation}
\label{MRIO3}
f(R)= \beta R^2 \ ,\quad
V(\varphi)= V_0 \left(\varphi - \varphi_c\right)\ .
\end{equation}
Here $\beta$ is a constant.  $R^2$ term is neglected by putting $\alpha_0=0$ in
(\ref{MR1}) since the curvature is small.
Searching for the solution (\ref{mGB2}) in sect.\ref{S2} and choosing $\varphi=\varphi_c +
\varphi_0/t^2$ or
$\varphi=\varphi_c + \varphi_0/\left(t_s -t\right)^2$, the following
restrictions are obtained \cite{inagaki}
\be
\label{MRIO12}
\varphi_0^2 = \frac{54\beta \left( -1 + 2h_0\right)^3
h_0^4}{\kappa^2\left(12h_0^2 - 2h_0 - 1\right)}\ ,\quad
V_0=\pm \frac{3h_0 + 1}{\sqrt{6\kappa^2\left(12h_0^2 - 2h_0 - 1\right)
\left(-1 + 2h_0\right)}}\ .
\ee
Since $\varphi_0^2$ should be positive, one finds
\begin{eqnarray}
\label{MRIO13}
& \mbox{when}\ \beta>0\ ,\quad
&\frac{1-\sqrt{13}}{12}<h_0<\frac{1+\sqrt{13}}{12}\
\mbox{or}\ h_0\geq \frac{1}{2}\ ,\nonumber \\
& \mbox{when}\ \beta<0\ ,\quad &h_0<\frac{1-\sqrt{13}}{12}\ \mbox{or}\
\frac{1+\sqrt{13}}{12}<h_0\leq \frac{1}{2}\ .
\end{eqnarray}
For example, if $h_0=-1/60$, which gives $w_{\rm eff}=- 1.025$, we find
\begin{equation}
\label{AA3}
\kappa V_0=\pm \frac{19}{34}\sqrt{\frac{15}{31}}=\pm 0.388722...\ .
\end{equation}
It is interesting that the value of $w_{\rm eff}$ is consistent with the
observed bounds. (Note that recently the above model has been reconsidered
in ref.\cite{S}).

For $h_0>0$ case, since $R=6\dot H + 12 H^2$,
the curvature $R$ decreases as $t^{-2}$ with time $t$ and $\varphi$
approaches to $\varphi_c$ but does not arrive at $\varphi_c$ in a finite time,
as expected in \cite{MR}.

As $H$ behaves as $h_0/t$ or $h_0/(t_s - t)$ for (\ref{mGB2}) in sect.\ref{S2}, if we
substitute the value of the age of the present universe $10^{10}$years$\sim
(10^{-33}$eV$)^{-1}$ into $t$ or $t_s-t$,
the observed value of $H$ could be reproduced, which could explain the
smallness of the effective cosmological constant $\Lambda\sim H^2$.
Note that even if there is no potential term, that is, $V_0=0$, when $\beta<0$,
there is a solution
\begin{equation}
\label{MRIO13b}
h_0=-\frac{1}{3}<\frac{1-\sqrt{13}}{12}=-0.2171...\ ,
\end{equation}
which gives the EoS parameter : $w=-3$,
although  $w$ is not realistic. Playing with different choices of the
potential and non-linear coupling more realistic predictions may be
obtained.

\subsection{Hierarchy problem in modified gravity \label{S43}}

   Hierarchy problem is known to be the fundamental one in
high energy physics. It is interesting to understand if modified gravity
approach may help in resolution of hierarchy problem?
Recently the hierarchy problem has been investigated, in \cite{BN},
in frames of scalar-tensor theory. Here, similar proposal will be
discussed in the modified gravity.
In \cite{BN}, in order to generate the hierarchy, a
small scale, which is the vacuum decay rate $\Gamma_{vac}$, was
considered. Instead of $\Gamma_{vac}$, we here use the age of the
universe $\sim 10^{-33}\,{\rm eV}$, as the small mass scale.

We now start from the action for modified gravity coupled
with matter (\ref{M1}) in sect.\ref{S2}.
Let us assume that the
matter Lagrangian density ${\cal L}_{\rm matter}$ contains a
Higgs-like scalar field $\varphi$
\be
\label{RR12bb}
{\cal L}_{\rm matter}= - \frac{1}{2}\nabla^\mu \varphi \nabla_\mu \varphi +
\frac{\mu^2}{2}\varphi^2 - \lambda \varphi^4 + \cdots \ .
\ee
Under the conformal transformation $g_{\mu\nu}\to \e^\sigma g_{\mu\nu}$, the matter
Lagrangian density ${\cal L}_{\rm matter}$ is transformed as
\be
\label{RR12a}
{\cal L}_{\rm matter} \to {\cal L}_{\rm matter}^\sigma
=  - \frac{\e^{\sigma}}{2}\nabla^\mu \varphi \nabla_\mu \varphi +
\frac{\mu^2\e^{2\sigma}}{2}\varphi^2 - \lambda \e^{2\sigma} \varphi^4 +\cdots \ .
\ee
By redefining $\varphi$ as $\varphi\to \e^{-\sigma/2}\varphi$,
${\cal L}_{\rm matter}^\sigma$ acquires the following form
\be
\label{RR12c}
{\cal L}_{\rm matter}^\sigma \sim  - \frac{1}{2}\nabla^\mu \varphi
\nabla_\mu \varphi + \frac{\mu^2\e^{\sigma}}{2}\varphi^2
   - \lambda \varphi^4 +\cdots \ ,
\ee
where  the time derivative of $\sigma$ is neglected. Then,
the massive parameter $\mu$, which determines the weak scale, is
effectively transformed as $\mu\to \tilde \mu\equiv \e^{\sigma/2}\mu$.
In principle, $\mu$ can be of the order of
the Planck scale $10^{19}$ GeV, but if $\e^{\sigma/2}\sim 10^{-17}$
in the present universe, $\tilde \mu$ could be of the order of $10^2$ GeV,
which is
the scale of the weak interaction. Therefore there is a quite
natural possibility that the hierarchy problem can be solved by
above version of modified gravity (similar solution of hierarchy problem
may be found in two scalar-tensor gravity of ref.\cite{ENO} as it has been
shown in \cite{cognola2}.)

One may consider the model \cite{cognola2}:
\be
\label{RR14b}
f(R) = R + f_0 R^\alpha\ ,
\ee
with constant $f_0$ and $\alpha$. (Without the first term the restrictions
to the parameter $\alpha$ have been recently studied in
ref.\cite{barrow2}.) If $\alpha<1$,
the second term dominates, when the curvature is small. Assuming
that the EoS parameter, $w$, of matter is constant one gets \cite{ANO}
the accelerated FRW cosmology solution with the effective EoS parameter $w_{\rm eff}$ (\ref{M9}).

Neglecting the first term in (\ref{RR14b}),
it follows
\be
\label{M10}
\e^{\sigma/2}\sim \frac{1}{\sqrt{f_0 \alpha R^{\alpha-1}}}\ .
\ee
In the present universe,  $R\sim
\left(10^{-33}\,{\rm eV}\right)^2$. Assume now that  $f_0$ is of the order
of the Planck scale $\sim 10^{19}\,{\rm GeV}=10^{28}\,{\rm
eV}$ as $f_0\sim \left(10^{28}\,{\rm
eV}\right)^{\frac{1}{2(\alpha-1)}}$. Then, Eq.~(\ref{M10})
yields
\be
\label{M11}
\e^{\sigma/2}\sim 10^{61\left(\alpha -1\right)}\ .
\ee
If we furthermore assume that $-17=61\left(\alpha
-1\right)$, we find $\alpha = 44/61$. In that case, if $w>-1$,
$w_{\rm eff}>-1$ and the universe is not in its phantom phase, but
(\ref{M11}) hints towards the possibility that modified
gravity can solve  the hierarchy problem.

Let us write the action of the scalar-tensor theory as
\bea
\label{F1}
S&=&{1 \over \kappa^2}\int d^4x
\sqrt{-g}\e^{\alpha\phi}\Bigl(R - {1 \over 2}
\partial_\mu \phi \partial^\mu \phi - V_0\e^{-2\phi/\phi_0})\Bigr)
+ \int d^4 x \sqrt{-g} {\cal L}_{\rm matter} \ , \nn
{\cal L}_{\rm matter} &=& - \frac{1}{2}\partial^\mu \varphi \partial_\mu \varphi
+ \frac{\mu^2}{2}\varphi^2 - \lambda \varphi^4 + \cdots \ .
\eea
This action could be regarded as the Jordan frame action (see sect.\ref{S2}).
Scalar field $\varphi$ could be identified with the Higgs field in the
weak(-electromagnetic) interaction.
Now the  ratio of the inverse of the the effective gravitational coupling
$\tilde\kappa=\kappa\e^{-{\alpha \phi \over 2}}$ and the Higgs mass  $\mu$ is given by
\be
\label{F2}
\frac{\frac{1}{\tilde\kappa}}{\mu}=\frac{\e^{{\alpha \phi \over 2}}}{\kappa\mu}\ .
\ee
Hence, even if both of $1/\kappa$ and $\mu$ are of the order of
the weak interaction scale,
if $\e^{{\alpha \phi \over 2}}\sim 10^{17}$, $1/\tilde\kappa$ could have
an order of the Planck scale.

By rescaling the metric and the Higgs scalar  $\varphi$ as
\be
\label{F3}
g_{\mu\nu}\to \e^{-\alpha \phi} g_{\mu\nu}\ ,\quad
\varphi \to \e^{{\alpha \phi \over 2}}\varphi\ ,
\ee
the Einstein frame action is obtained:
\bea
\label{F4}
S&=&{1 \over \kappa^2}\int d^4x
\sqrt{-g}\Bigl(R - {1 \over 2}\left(1 + 3\alpha^2\right)
\partial_\mu \phi \partial^\mu \phi - V_0\e^{-2\phi\left(1/\phi_0+\alpha\right)})\Bigr) \nn
&& + \int d^4 x \sqrt{-g} {\cal L}_{\rm matter} \ , \nn
{\cal L}_{\rm matter} &=& - \frac{1}{2}
\left(\partial^\mu \varphi + \frac{\alpha}{2}\partial^\mu\phi \varphi\right)
\left(\partial_\mu \varphi + \frac{\alpha}{2}\partial_\mu\phi \varphi\right)
+ \frac{\mu^2\e^{-\alpha \phi}}{2}\varphi^2 - \lambda \varphi^4 + \cdots \ .
\eea
In the Einstein frame, the gravitational coupling $\kappa$ is constant
   but the effective Higgs mass
$\tilde \mu$ defined by $\tilde\mu \equiv \e^{-{\alpha \phi \over 2}}\mu$
can be time-dependent.
Hence, the ratio of $1/\kappa$ and $\tilde\mu$ is given by
\be
\label{F6}
\frac{\frac{1}{\kappa}}{\tilde\mu}=\frac{\e^{{\alpha \phi \over 2}}}{\kappa\mu}\ ,
\ee
which is identical with (\ref{F2}).
Hence, even if both of $1/\kappa$ and $\mu$ are of the order of the Planck
scale,
if $\e^{{\alpha \phi \over 2}}\sim 10^{17}$, $\tilde\mu$ could have  an
order
of  the weak interaction scale.
Therefore the solution of the hierarchy problem does not essentially
depend on the frame choice.

Nevertheless, note that the cosmological time variables in two frames
could be different due to the
scale transformation (\ref{F3}) as $dt \to d\tilde t = \e^{-{\alpha \phi \over 2}}dt$.
Therefore the time-intervals are different in the two frames.
The units of time and the length are
now defined by the electromagnetism. Then the frame where the electromagnetic fields do not
couple with the scalar field $\phi$ could be physically more preferrable.
Since the electromagnetic interaction is a part
of the electro-weak interaction, the Jordan frame in (\ref{F1}) could be
more preferrable from
the point of view of the solution of hierarchy problem.

\section{Late-time cosmology in modified Gauss-Bonnet gravity \label{S5}}

\subsection{$f(G)$ gravity \label{S51}}

Our next example is modified Gauss-Bonnet gravity.
Let us start from the action \cite{GB,cognola2}:
\be
\label{GB1b}
S=\int d^4x\sqrt{-g} \left(\frac{1}{2\kappa^2}R + f(G) + {\cal L}_m\right)\ .
\ee
Here
${\cal L}_m$ is the matter Lagrangian density and $G$ is the GB
invariant: $G=R^2 -4 R_{\mu\nu} R^{\mu\nu} + R_{\mu\nu\xi\sigma}
R^{\mu\nu\xi\sigma}$. By variation over $g_{\mu\nu}$
one gets:
\bea
\label{GB4b} && 0= \frac{1}{2\kappa^2}\left(- R^{\mu\nu} +
\frac{1}{2} g^{\mu\nu} R\right) + T^{\mu\nu} + \frac{1}{2}g^{\mu\nu} f(G)
   -2 f'(G) R R^{\mu\nu} \nn
&& + 4f'(G)R^\mu_{\ \rho} R^{\nu\rho}
   -2 f'(G) R^{\mu\rho\sigma\tau}R^\nu_{\ \rho\sigma\tau}
   -4 f'(G) R^{\mu\rho\sigma\nu}R_{\rho\sigma}
+ 2 \left( \nabla^\mu \nabla^\nu f'(G)\right)R \nn
&& - 2 g^{\mu\nu} \left( \nabla^2 f'(G)\right)R
   - 4 \left( \nabla_\rho \nabla^\mu f'(G)\right)R^{\nu\rho}
   - 4 \left( \nabla_\rho \nabla^\nu f'(G)\right)R^{\mu\rho} \nn
&& + 4 \left( \nabla^2 f'(G) \right)R^{\mu\nu} + 4g^{\mu\nu} \left(
\nabla_{\rho} \nabla_\sigma f'(G) \right) R^{\rho\sigma}
   - 4 \left(\nabla_\rho \nabla_\sigma f'(G) \right) R^{\mu\rho\nu\sigma} \ ,
\eea
where $T^{\mu\nu}$ is the matter EM tensor.
By choosing the spatially-flat FRW universe metric (\ref{RR33}) in sect.\ref{S2},
the equation corresponding to the first FRW equation has the following form:
\be
\label{GB7b}
0=-\frac{3}{\kappa^2}H^2 + Gf'(G) - f(G)
   - 24 \dot Gf''(G) H^3 + \rho_m\ ,
\ee
where $\rho_m$ is the matter energy density.
When $\rho_m=0$, Eq.~(\ref{GB7b}) has a deSitter universe solution where
$H$, and therefore $G$, are constant. For $H=H_0$, with constant
$H_0$, Eq.~(\ref{GB7b}) turns into
\be
\label{GB7bb}
0=-\frac{3}{\kappa^2}H_0^2 + 24H_0^4 f'\left( 24H_0^4 \right) -
f\left( 24H_0^4\right) \ .
\ee
For a large number of choices of the
function $f(G)$, Eq.~(\ref{GB7bb}) has a non-trivial ($H_0\neq 0$)
real solution for $H_0$ (deSitter universe). The late-time cosmology for
above theory without matter has been discussed for a number of examples
in refs.\cite{GB,cognola2}. We follow these works here.

We now consider the case $\rho_m\neq 0$. Assuming that the
EoS parameter $w\equiv p_m/\rho_m$ for matter ($p_m$ is the
pressure of matter) is a constant then, by using the conservation of
energy: $\dot \rho_m + 3H\left(\rho_m + p_m\right)=0$, we find
$\rho=\rho_0 a^{-3(1+w)}$. The function $f(G)$ is chosen as
\be
\label{mGB1}
f(G)=f_0\left|G\right|^\beta\ ,
\ee
with constant
$f_0$ and $\beta$. If $\beta<1/2$, $f(G)$ term becomes dominant
compared with the Einstein term when the curvature is small. If we
neglect the contribution from the Einstein term in (\ref{GB7b}),
assuming FRW anzats (\ref{mGB2}) in sect.\ref{S2}, the following solution may be found
\be
\label{mGB3}
h_0=\frac{4\beta}{3(1+w)}\ ,\quad
a_0=\Bigl[ -\frac{f_0(\beta - 1)}
{\left(h_0 - 1\right)\rho_0}\left\{24 \left|h_0^3 \left(- 1 +
h_0\right) \right|\right\}^\beta \left( h_0 - 1 +
4\beta\right)\Bigr]^{-\frac{1}{3(1+w)}}\ .
\ee
Then the effective EoS parameter $w_{\rm eff}$ (\ref{FRW3k})  in sect.\ref{S2}
is less than $-1$ if $\beta<0$, and for $w>-1$ is
\be
\label{mGB3b}
w_{\rm eff}=-1 + \frac{2}{3h_0}=-1 +
\frac{1+w}{2\beta}\ ,
\ee
which is again less than $-1$ for
$\beta<0$. Thus, if $\beta<0$, we obtain an effective phantom with
negative $h_0$ even in the case when $w>-1$. In the phantom phase,
there might seem to occur the Big Rip  at $t=t_s$ \cite{brett}.
Near this Big Rip , however, the curvature becomes
dominant and then the Einstein term dominates, so that the
$f(G)$ term can be neglected. Therefore, the universe behaves as
$a=a_0t^{2/3(w+1)}$ and as a consequence the Big Rip does not
  eventually occur. The phantom era is transient.

The case when $0<\beta<1/2$ may be also considered. As $\beta$ is
positive, the universe does not reach here the phantom phase. When
the curvature is strong, the $f(G)$ term in the action (\ref{GB1b})
can be neglected and we can work with Einstein gravity. Then if
$w$ is positive, the matter energy density $\rho_m$ should behave as
$\rho_m\sim t^{-2}$, but $f(G)$ goes as $f(G)\sim t^{-4\beta}$.
Then, for late times (large $t$), the $f(G)$ term may become
dominant as compared with the matter one. If we neglect the
contribution from matter, Eq.~(\ref{GB7b}) has a deSitter universe
solution where $H$, and therefore $G$, are constant. If $H=H_0$ with
constant $H_0$, Eq.~(\ref{GB7b}) looks as (\ref{GB7bb}). As a
consequence, even if we start from the deceleration phase with
$w>-1/3$, we may also reach an asymptotically deSitter universe,
which is an accelerated universe. Correspondingly, also here there
could be a transition from acceleration to deceleration of the
universe.

Now, we consider the case when the contributions coming from the
Einstein and matter terms can be neglected. Then, Eq.~(\ref{GB7b})
reduces to
\be
\label{mGB9}
0=Gf'(G) - f(G) - 24 \dot Gf''(G) H^3 \ .
\ee
If $f(G)$ behaves as (\ref{mGB1}), from assumption
(\ref{mGB2}) in sect.\ref{S2}, we obtain
\be
\label{mGB10}
0=\left(\beta -
1\right)h_0^6\left(h_0 - 1\right) \left(h_0 - 1 + 4\beta \right)\ .
\ee
As $h_0=1$ implies $G=0$, one may choose
\be
\label{mGB11}
h_0 = 1 - 4\beta\ ,
\ee
and Eq.~(\ref{FRW3k}) in sect.\ref{S2} gives
\be
\label{mGB12}
w_{\rm eff}=-1 + \frac{2}{3(1-4\beta)}\ .
\ee
Therefore, if
$\beta>0$, the universe is accelerating ($w_{\rm eff}<-1/3$) and if
$\beta>1/4$, the universe is in a phantom phase ($w_{\rm eff}<-1$).
Thus, we are led to consider the following model:
\be
\label{mGB13}
f(G)=f_i\left|G\right|^{\beta_i} + f_l\left|G\right|^{\beta_l} \ ,
\ee
where we assume that
\be
\label{mGB14}
\beta_i>\frac{1}{2}\ ,\quad \frac{1}{2}>\beta_l>\frac{1}{4}\ .
\ee
Then, when the
curvature is large, as in the primordial universe, the first term
dominates, compared with the second one and the Einstein term, and
gives
\be
\label{mGB15}
-1>w_{\rm eff}=-1 +
\frac{2}{3(1-4\beta_i)}>-5/3\ .
\ee
On the other hand, when the
curvature is small, as is the case in the present universe, the
second term in (\ref{mGB13}) dominates, compared with the first one
and the Einstein term, and yields
\be
\label{mGB16} w_{\rm eff}=-1 +
\frac{2}{3(1-4\beta_l)}<-5/3\ .
\ee
Therefore, theory (\ref{mGB13}) can in
fact produce a model which is able to describe inflation and the
late-time acceleration of the universe in the unified way.

Instead of (\ref{mGB14}), one may also choose $\beta_l$ as
\be
\label{mGB17}
\frac{1}{4}>\beta_l>0\ ,
\ee
which gives
\be
\label{mGB18} -\frac{1}{3}>w_{\rm eff}>-1\ .
\ee
Then, what we obtain is an effective quintessence. Moreover, by properly adjusting
the couplings $f_i$ and $f_l$ in (\ref{mGB13}), one can obtain a
period where the Einstein term dominates and the universe is in a
deceleration phase. After that, there would come a transition from
deceleration to acceleration, when the GB term becomes the dominant
one. More choices of $f(G)$ may be studied with the purpose of
the construction of natural accelerating universes.

Let us address the issue of the correction to Newton law. Let
$g_{(0)}$ be a solution of (\ref{GB4b}) and represent the
perturbation of the metric as $g_{\mu\nu}=g_{(0)\mu\nu} +
h_{\mu\nu}$. First, we consider the perturbation around the deSitter
background which is a solution of (\ref{GB7bb}). We write  the
deSitter space metric as $g_{(0)\mu\nu}$, which gives the following
Riemann tensor:
\be
\label{GB35}
R_{(0)\mu\nu\rho\sigma}=H_0^2\left(g_{(0)\mu\rho}g_{(0)\nu\sigma}
   - g_{(0)\mu\sigma}g_{(0)\nu\rho}\right)\ .
\ee
The flat background corresponds to the limit of $H_0\to 0$. For
simplicity,  the following gauge condition is chosen:
$g_{(0)}^{\mu\nu} h_{\mu\nu}=\nabla_{(0)}^\mu h_{\mu\nu}=0$. Then
Eq.~(\ref{GB4b}) gives
\be
\label{GB38b}
0=\frac{1}{4\kappa^2} \left(\nabla^2 h_{\mu\nu}
   - 2H_0^2 h_{\mu\nu}\right) + T_{\mu\nu}\ .
\ee
The GB term contribution  does not appear except
in the length parameter $1/H_0$ of the deSitter space, which is
determined with account to the GB term.
This may occur due to the special structure of  GB invariant.
Eq.~(\ref{GB38b}) tells us
that there is no correction to Newton's law in deSitter and even in
the flat background corresponding to $H_0\to 0$, whatever is the
form of $f$ (at least, with above gauge condition). Note that
the study of newtonian limit in $1/R$ gravity (where significant
corrections to newton law may appear) and its extensions
has been done in \cite{newton,SNPRD}. For most $1/R$ models
the corrections to Newton law do not comply with solar system tests.

By introducing two auxilliary fields, $A$ and $B$, one can rewrite
the action (\ref{GB1b}) as
\bea
\label{GB3}
S&=&\int d^4 x\sqrt{-g}\Bigl(\frac{1}{2\kappa^2}R + B\left(G-A\right) \nn
&& + f(A) + {\cal L}_m \Bigr)\ .
\eea
Varying over $B$, it follows that
$A=G$. Using this in (\ref{GB3}), the action  (\ref{GB1b}) is recovered.
On the other hand, varying over $A$ in (\ref{GB3}), one gets $B=f'(A)$, and hence
\be
\label{GB6}
S=\int d^4
x\sqrt{-g}\Bigl(\frac{1}{2\kappa^2}R + f'(A)G - Af'(A) + f(A)\Bigr)\ .
\ee
By varying over $A$, the relation $A=G$ is
obtained again. The scalar is not dynamical as it has no kinetic
term. We may add, however, a kinetic term to the action by hand
\be
\label{GB6b}
S=\int d^4 x\sqrt{-g}\Bigl(\frac{1}{2\kappa^2}R -
\frac{\epsilon}{2}\partial_\mu A \partial^\mu A + f'(A)G - Af'(A) + f(A)\Bigr)\ .
\ee
Then one obtains a dynamical scalar theory
coupled with the Gauss-Bonnet invariant and with a potential. It is
known that a theory of this kind  has no ghosts and it is stable, in
general. Actually, it is related with string-inspired dilaton gravity
proposed as alternative for dark energy \cite{NOS,stringT}.
Then, in case that the limit $\epsilon\to 0$ can be
obtained smoothly, the corresponding $f(G)$ theory has no
ghost and could actually be stable. It may be of interest to study the
cosmology of such theory in the limit $\epsilon\to 0$.

\subsection{$f(G,R)$ gravity \label{S52}}

It is interesting to study late-time cosmology in generalized
theories, which include both the functional dependence from
curvature as well as from the Gauss-Bonnet term \cite{cognola2}:
\be
\label{GR1b}
S=\int d^4 x\sqrt{-g}\left(f(G,R) + {\cal L}_m\right)\ .
\ee
The following solvable model is considered:
\be
\label{GR7b}
f(G,R)=R \tilde f\left(\frac{G}{R^2}\right)\ ,
\quad \tilde f\left(\frac{G}{R^2}\right)=\frac{1}{2\kappa^2}
+ f_0 \left(\frac{G}{R^2}\right)\ .
\ee
The FRW solution may be found again:
\be
\label{GR11}
H=\frac{h_0}{t}\ ,\quad
h_0 = \frac{\frac{3}{\kappa^2}
   - 2f_0 \pm \sqrt{8f_0\left(f_0 - \frac{3}{8\kappa^2}\right)}}
{\frac{6}{\kappa^2} + 2f_0}\ .
\ee
Then, for example, if $\kappa^2 f_0<-3$, there is a solution
describing a phantom with $h_0<-1-\sqrt{2}$ and a solution
describing the effective matter with $h_0>-1+\sqrt{2}$.
Late-time cosmology in other versions of such theory may be constructed.

\section{Inhomogeneous equation of  state of the universe dark fluid \label{S6}}

Let us remind several simple facts about the universe filled with
ideal fluid.
By using the energy conservation law $0=\dot\rho + 3H\left(p + \rho\right)$,
when $\rho$ and $p$ satisfy the following simple EOS $p=w\rho$ with constant $w$,
we find $\rho=\rho_0 a^{-3(1+w)}$. Then by using the first  FRW equation
$(3/\kappa^2)H^2=\rho$, the well-known solution follows
$a=a_0 \left(t - t_1\right)^\frac{2}{3(w+1)}$ $\left(w>-1\right)$ or
$a_0 \left(t_2 - t\right)^\frac{2}{3(w+1)}$ $w\neq -1$ $\left(w<-1\right)$ and
$a=a_0\e^{\kappa t\sqrt{\frac{\rho_0}{3}}}$ when $w=-1$, which is the deSitter universe.
Here $t_1$ and $t_2$ are constants of the integration.
When $w<-1$, there appears a Big Rip singularity in a finite
time at $t=t_2$.

In general, the singularities in dark energy universe may behave in a
different way.
One may classify the future singularities as follows \cite{tsujikawa}:
\begin{itemize}
\item  Type I (``Big Rip'') : For $t \to t_s$, $a \to \infty$,
$\rho \to \infty$ and $|p| \to \infty$
\item  Type II (``sudden'') : For $t \to t_s$, $a \to a_s$,
$\rho \to \rho_s$ or $0$ and $|p| \to \infty$
\item  Type III : For $t \to t_s$, $a \to a_s$,
$\rho \to \infty$ and $|p| \to \infty$
\item  Type IV : For $t \to t_s$, $a \to a_s$,
$\rho \to 0$, $|p| \to 0$ and higher derivatives of $H$ diverge.
This also includes the case when $\rho$ ($p$) or both of them
tend to some finite values while higher derivatives of $H$ diverge.
\end{itemize}
Here $t_s$, $a_s$ and $\rho_s$ are constants with $a_s\neq 0$.
The type I may correspond to the Big Rip singularity,
which emerges when constant $w$ is less than $-1$.
The type II corresponds to the sudden future
singularity \cite{Barrow} at which $a$ and $\rho$ are finite but $p$ diverges.
The type III appears for the model with $p=-\rho-A \rho^{\alpha}$
\cite{final,Stefancic}, which is different from the sudden future
singularity
in the sense that $\rho$ diverges.
Now, let us discuss the universe filled with ideal fluid with
inhomogeneous EoS. (It is assumed that GR is valid). Effectively, this
approach overlaps with modified
gravity one as inhomogeneous contribution may be considered as one coming
from new gravitational terms. Moreover, it may be an extremely useful in
the situation when modified gravity represents highly non-linear theory
where only some features (but not the whole Lagrangian) are known
like is currently the case with M-theory.

\subsection{The singularities in the inhomogeneous EoS dark fluid
universe \label{S61}}

One may start from the dark fluid with the following EOS:
\be
\label{EOS1}
p=-\rho - f(\rho)\ ,
\ee
where $f(\rho)$ can be an arbitrary function in general.
The choice $f(\rho)\propto \rho^\alpha$ with a constant $\alpha$
was proposed in Ref.\cite{final} and was investigated in detail in
Ref.\cite{Stefancic,tsujikawa}.
Then the scale factor is given by
\be
\label{EOS4}
a=a_0\exp\left(\frac{1}{3} \int \frac{d\rho}{f(\rho )} \right)\ ,
\ee
and the cosmological time may be found
\be
\label{tint}
t=\int \frac{d \rho}{\kappa \sqrt{3\rho} f(\rho)}\ ,
\ee

As an example we may consider the case that
\be
\label{ppH6}
f(\rho)=A\rho^\alpha\ .
\ee
Then we find \cite{tsujikawa}:
\begin{itemize}
\item In case $\alpha=1/2$ or $\alpha=0$, there does not appear any
singularity.
\item In case $\alpha>1$, when $t\to t_0$, the energy density behaves
as $\rho\to\infty$ and therefore $|p|\to \infty$.
Then the scale factor $a$ is finite even if $\rho\to \infty$. Therefore
$\alpha>1$ case
corresponds to type III singularity.
\item In $\alpha=1$ case, if $A>0$, there occurs the Big Rip or type I singularity
but if $A\leq 0$, there does not appear future singularity.
\item In case $1/2<\alpha<1$, when $t\to t_0$, all of $\rho$, $|p|$, and $a$
diverge if $A>0$
then this corresponds to type I singularity.
\item In case $0<\alpha<1/2$, when $t\to t_0$, we find $\rho$, $|p|\to 0$ and
$a\to a_0$ but
\be
\label{typeIV}
\ln a \sim \left|t-t_0\right|^{\frac{\alpha-1}{\alpha - 1/2}}\ .
\ee
Since the exponent $(\alpha -1)/(\alpha - 1/2)$ is not always an integer, even if $a$
is finite, the higher derivatives of $H$ diverge in general. Therefore this
case corresponds to type IV singularity.
\item In case $\alpha<0$, when $t\to t_0$, we find $\rho\to 0$, $a\to a_0$ but $|p|\to \infty$.
Therefore this case corresponds to type II singularity.
\end{itemize}
Hence, the brief review of dark fluid FRW cosmology with specific
homogeneous EOS as well
as its late-time behaviour (singularities) is given (see
\cite{tsujikawa} for more
detail).

At the next step, we  consider the inhomogeneous EOS for dark fluid,
so that the dependence from Hubble parameter is included in EOS.
This new terms may origin from string/M-theory, braneworld or modified
gravity.
One more motivation for such EOS comes from including of time-dependent
bulk
viscosity in dark fluid EOS \cite{BG} or from the modification of gravity.
Hence, we suggest the following EOS \cite{inh}
\be
\label{pH1}
p=-\rho + f(\rho) + G(H)\ .
\ee
where $G(H)$ is some function.

As an example, one may consider the case
\be
\label{ppH9}
f(\rho)=A\rho^\alpha \to f(\rho) + G(H) = - A\rho^{\alpha} - B H^{2\beta}\ .
\ee
Here, it follows \cite{inh}

\begin{itemize}
\item In case $\alpha>1$, for most values of $\beta$, there occurs type III
singularity.
In addition to the type III singularity, when $0<\beta<1/2$, there  occurs type IV
singularity and when $\beta<0$, there  occurs type II singularity.
\item $\alpha=1$ case, if $\beta>1$, the singularity becomes type III.
If $\beta<1$ and $A>0$, there
occurs the Big Rip or type I singularity. In addition to the type I singularity,
we  have type IV singularity when $0<\beta<1/2$ and type II when $\beta<1$.
\item In case $1/2<\alpha<1$, one sees singularity of type III for $\beta>1$,
type I for $1/2\leq\beta<1$ (even for $\beta=1/2$) or $\beta=1$ and $B'>0$ ($B>0$)
case. In addition to type I, type IV case occurs for $0<\beta<1/2$, and type II
for $\beta<0$.
\item In case $\alpha=1/2$, we have singularity of type III for $\beta>1$, type
I for $1/2<\beta<1$ or $\beta=1$ and $B'>0$ ($B>0$), type IV for $0<\beta<1/2$,
and type II for $\beta<0$.
When $\beta=1/2$ or $\beta=0$, there does not appear any singularity.
\item In case $0<\alpha<1/2$, we find type IV for $0<\beta<1/2$, and type II
for $\beta<0$. In addition to type IV singularity,  there occurs singularity of
type III for $\beta>1$, type I for $1/2\leq\beta<1$ or $\beta=1$ and $B'>0$ ($B>0$)  case.
\item In case $\alpha<0$, there will always occur type II singularity.
In addition to type II singularity, we  have a singularity of
type III for $\beta>1$, type I for $1/2\leq\beta<1$ or $\beta=1$ and $B'>0$ ($B>0$)  case.
\end{itemize}

Thus, we demonstrated how the modification of EOS by Hubble dependent,
inhomogeneous term changes the structure of singularities in late-time
dark
energy universe.

In general, EOS needs to be double-valued in order for
the transition (crossing of phantom divide) to occur between the region
$w<-1$ and the region $w>-1$.
Then there could not be one-to-one correspondence between $p$ and $\rho$.
In such a case, instead of (\ref{pH1}), we may suggest the implicit,
inhomogeneous equation of the state
\be
\label{pH11}
F(p,\rho,H)=0\ .
\ee
The following example may be of interest:
\be
\label{pH12}
\left(p+\rho\right)^2 - C_0 \rho^2 \left(1 - \frac{H_0}{H}\right)=0\ .
\ee
Here $C_0$ and $H_0$ are positive constants.
Hence, the Hubble rate looks as
\be
\label{pH14}
H=\frac{16}{9C_0^2 H_0 \left(t-t_-\right)\left(t_+ - t\right)}\ .
\ee
and
\be
\label{pH16}
p=-\rho\left\{1+\frac{3C_0^2}{4H_0}\left(t-t_0\right)\right\}\ ,\quad
\rho =\frac{2^8}{3^3C_0^4 H_0^2 \kappa^2 \left(t-t_-\right)^2
\left(t_+ - t\right)^2}\ .
\ee
In (\ref{pH14}), since $t_-<t_0<t_+$, as long as $t_-<t<t_+$, the Hubble rate
$H$ is positive.
The Hubble rate $H$ has a minimum $H=H_0$ when $t=t_0=\left(t_- + t_+\right)/2$
and diverges when $t\to t_\pm$.
Then one may regard $t\to t_-$ as a Big Bang singularity and $t\to t_+$ as
a Big Rip one.
As clear from (\ref{pH16}), the parameter $w=p/\rho$ is larger than $-1$ when $t_-<t<t_0$
and smaller than $-1$ when $t_0<t<t_+$. Therefore there occurs the
crossing of
phantom divide $w=-1$ when $t=t_0$ thanks to the effect of inhomogeneous term in EOS.

In principle, the more general EOS may contain the derivatives of $H$, like
$\dot H$, $\ddot H$, ... More general EOS than (\ref{pH11}) may have the
following form:
\be
\label{dH1}
F\left(p,\rho,H,\dot H,\ddot H,\cdots\right)=0\ .
\ee
Trivial example is that
\be
\label{dH2}
p=w\rho - \frac{2}{\kappa^2}\dot H - \frac{3(1+w)}{\kappa^2}H^2\ .
\ee
By using the first and second FRW equations
$\rho=(3/\kappa^2)H^2$ and $p=-(2/\kappa^2)\dot H - (3/\kappa^2)H^2$,
Eq.(\ref{dH2}) becomes an identity, which means that any FRW cosmology
can be a solution if EOS (\ref{dH2}) is assumed.

\subsection{Origin of the inhomogeneous terms \label{S62}}

In order to give the support for an inhomogeneous EoS dark fluid \cite{CNO}
we will indicate how the extra terms may appear from the modifications of
General Relativity.
Let us start with the action of the scalar-tensor theory (\ref{k1}) in
sect.\ref{S2}.
The energy density $\rho$
and the pressure $p$ for the scalar field $\phi$ are given by
\be
\label{k4}
\rho = \frac{1}{2}\omega(\phi){\dot \phi}^2 + V(\phi)\ ,
\quad p = \frac{1}{2}\omega(\phi){\dot \phi}^2 - V(\phi)\ .
\ee
Since we can always redefine the scalar field $\phi$ as $\phi\to F(\phi)$ by an arbitrary function
$F(\phi)$, we can choose
the scalar field to be a time coordinate; $\phi=t$. Furthermore we consider the case that
$\omega(\phi)$ and $V(\phi)$ are given by a single function $f(\phi)$ as
(see sect.\ref{S24})
\be
\label{any5}
\omega(\phi) = - \frac{2}{\kappa^2}f'(\phi) \ ,\quad
V(\phi) = \frac{1}{\kappa^2}\left(3f(\phi)^2 + f'(\phi)\right) \ .
\ee
Hence, one finds
\be
\label{kk1}
\rho= \frac{3}{\kappa^2}f(\phi)^2\ ,\quad
p= -\frac{3}{\kappa^2}f(\phi)^2 - \frac{2}{\kappa^2}f'(\phi)\ .
\ee
Since $\rho=f^{-1}\left(\kappa\sqrt{\rho/3}\right)$, we obtain the
inhomogeneous
equation of the state
\be
\label{SN1}
p=-\rho -
\frac{2}{\kappa^2}f'\left(f^{-1}\left(\kappa\sqrt{\frac{\rho}{3}}\right)\right)\ .
\ee
As an example, we consider the case
\be
\label{SN2}
f=f_0\phi^2 + f_1\ ,
\ee
with constant $f_0$ and $f_1$.
Then (\ref{SN1}) gives
\be
\label{SN3}
0=\left(p + \rho\right)^2 - \frac{4f_0}{\kappa^4} f\left(1 - \frac{f_1}{f}\right)\ .
\ee
If the matter contribution is neglected, from the first FRW equation, one
finds
\be
\label{kk2}
f^2=\frac{\kappa^2}{3}\rho =H^2\ .
\ee
Then we may rewrite (\ref{SN3}) as
\be
\label{kk3}
0=\left(p + \rho\right)^2 - \frac{4f_0}{\kappa^3} \sqrt{\frac{\rho}{3}}
\left(1 - \frac{f_1}{H}\right)\ ,
\ee
which  has a structure similar to (\ref{pH12}).
One may also consider the case that
\be
\label{kkk1}
f=\frac{f_0}{\phi}\ ,
\ee
which gives
\be
\label{kkk2}
p=\left(-1 + \frac{2}{3f_0}\right)\rho\ .
\ee
Then we obtain dark fluid with homogeneous EoS with $w=-1 +
{2}/{3f_0}$. Note that despite the fact that specific scalar-tensor theory
and dark fluid may predict the same FRW dynamics, their physical
properties may be different. For instance, Newton law, number of FRW
solutions, their stability properties may significally differ
in two formulations.

Another possibility to obtain the  inhomogeneous EOS dark fluid is related
  with the modified  gravity.
As an illustrative example,  the following action is considered:
\be
\label{HD1b}
S=\int d^4 x \sqrt{-g}\left(\frac{1}{2\kappa^2}R + {\cal L}_{\rm matter}
+ f(R)\right)\ .
\ee

In the FRW universe, the gravitational
equations are:
\bea
\label{HD2b}
0&=& - \frac{3}{\kappa^2}H^2 + \rho - f\left(R=6\dot H + 12 H^2\right) \nn
&& + 6\left(\dot H + H^2 - H \frac{d}{dt}\right) f'\left(R=6\dot H + 12 H^2\right) \ ,\\
\label{HD3b}
0&=& \frac{1}{\kappa^2}\left(2\dot H + 3H^2\right) + p + f\left(R=6\dot H + 12 H^2\right) \nn
&& + 2\left( - \dot H - 3H^2 + \frac{d^2}{dt^2} + 2H \frac{d}{dt}\right)
f'\left(R=6\dot H + 12 H^2\right)\ .
\eea
Here $\rho$ and $p$ are the energy density and the pressure coming from ${\cal
L}_{\rm matter}$.
They may satisfy the equation of state like $p=w\rho$.
One may now define the effective energy density $\tilde \rho$ and $\tilde p$ by
\bea
\label{HD4}
\tilde\rho &\equiv& \rho - f\left(R=6\dot H + 12 H^2\right)
+ 6\left(\dot H + H^2 - H \frac{d}{dt}\right) f'\left(R=6\dot H + 12 H^2\right) \ ,\\
\label{HD5}
\tilde p&=& p + f\left(R=6\dot H + 12 H^2\right) \nn
&& + 2\left( - \dot H - 3H^2 + \frac{d^2}{dt^2} + 2H \frac{d}{dt}\right)
f'\left(R=6\dot H + 12 H^2\right)\ .
\eea
Thus, it follows
\bea
\label{HD6}
\tilde p &=& w\tilde \rho  + (1+w)f\left(R=6\dot H + 12 H^2\right)
+ 2\left( \left(-1 - 3w\right) \dot H - 3\left(1+w\right) H^2
+ \frac{d^2}{dt^2} \right. \nn
&& \left. + \left(2 + 3w\right) H \frac{d}{dt}\right)
f'\left(R=6\dot H + 12 H^2\right) \ .
\eea
Especially if we consider the case that $f=f_0R$ with a constant $f_0$, we
obtain
\be
\label{kk4}
\tilde p= w {\tilde \rho} + 6f_0\dot H + 6(1+w)f_0 H^2\ .
\ee
Thus, FRW dynamics (here, the same FRW equations) of some modified gravity
may be equivalent to FRW
dynamics in GR coupled with the inhomogeneous EoS dark fluid.
Of course, the physics of two equivalent FRW descriptions (Newton law,
perturbations structure, stability of FRW solution, number of FRW
solutions,etc) may be quite
different \cite{e}. In this respect, one should note
that different modified gravities may also lead to the same FRW equations
(see an explicit example in \cite{multamaki}). Nevertheless,
the details of FRW dynamics will be again different as modified gravity
predictions should be compared with observations (for some attempts in
this direction,
see \cite{CNO,obs,KK}).

\section{Discussion \label{S7}}

In summary, we presented the introduction to some classes of
modified gravities considered as gravitational alternative for dark
energy. Specifically, the models which include $f(R)$, $f(G)$, $f(R,G)$
terms or string-inspired GB-dilaton terms are considered (for recent
review of higher derivative gravities, see \cite{hjs2} and for
discussion of their experimental manifestations, see \cite{rizzo}).
Special attention is paid to their versions where the action grows with
decrease
of the scalar curvature.
In addition, the FRW universe filled with dark fluid with inhomogeneous EoS
which may be induced by modified gravity is discussed.
As it has been stressed in the introduction and as it has been explicitly
demonstrated above,
modified gravity is endowed with a quite rich cosmological structure:
it may naturally lead to an effective cosmological constant, quintessence
or phantom era in the late universe with a possible transition from
deceleration to acceleration or (if necessary) the crossing of phantom
divide. Moreover, some versions of modified gravity
may pass the Solar System tests quite successfully.

Nevertheless, we are still missing the modified gravity theory
which, from one side, is as good as General Relativity in Solar System
tests and from another side contains the fully consistent
explanation for current cosmic acceleration.
(Note that if universe is currently entering to phantom phase, then some deviations
from General Relativity are quite expectable at large distances). It is
quite possible that
other versions of modified gravity: string-inspired gravity with
fourth-order (or higher) in $R$ corrections, theories which contain Traces
of inverse
Riemann tensor \cite{ESTW} or negative powers of squared Riemann and Ricci
tensors \cite{turner} or even non-local Riemann  and Ricci tensor
terms
(for instance, of the sort
induced by Quantum Field Theory \cite{eli} with terms like
$R/R_{\mu\nu}\Box R^{\mu\nu}$, $R_{\mu\nu} R^{\mu\nu}/R\Box R$, etc)
or some combination of the above models with the ones under discussion
should be considered as dark energy candidates as well.
  From another point, the next generation of precise observations
should define the evolution of the cosmological parameters,
and, perhaps, to rule out some of possible gravitational dark energy
models leading to viable explanation of dark energy problem.

\section*{Acknowledgments}

We would like to thank G. Allemandi, I. Brevik, A. Borowiec,
S. Capozziello, G. Cognola, E. Elizalde, M. Francaviglia,
S. Tsujikawa, M. Sasaki, M. Sami, L. Vanzo and S. Zerbini for helpful
discussions on related questions.
The research by SDO has been supported in part by LRSS project
n4489.2006.02 (RFBR, Russia), by the project
FIS2005-01181 (MEC, Spain) and by the project 2005SGR00790 (AGAUR,
Catalunya, Spain).

\end{document}